\documentclass[12pt]{article}

\usepackage{amsmath,amsfonts}
\usepackage{amssymb}
\usepackage{hhline}
\usepackage{epsfig,cite}
\usepackage{stmaryrd}
\usepackage[usenames,dvips]{color}
\usepackage{fullpage}
\usepackage{verbatim}
   \allowdisplaybreaks
\makeatletter
\@addtoreset{equation}{section}
\makeatother

\newcommand{\be}{\begin{equation}}
\newcommand{\ee}{\end{equation}}
\newcommand{\bea}{\begin{eqnarray}}
\newcommand{\eea}{\end{eqnarray}}

\newcommand{\diag}{\operatorname{diag}}

\newcommand{\nn}{\nonumber}

\newcommand{\wv}{\psi}
\newcommand{\su}{{\cal W}}
\newcommand{\Op}{\hat{\mathcal{O}}}

\begin{document}

\thispagestyle{empty}

\begin{center}
\hfill UAB-FT-743
\begin{center}

\vspace{.5cm}

{\Large\sc
Unitarity and singular backgrounds}

\end{center}

\vspace{1.cm}

\textbf{ Nikos Brouzakis$^{\,a}$, and Mariano Quiros$^{\,b}$}\\

\vspace{1.cm}
${}^a\!\!$ {\em {Department of Physics, University of Athens, University Campus \\Zographou 15784, Greece}}

\vspace{.1cm}

${}^b\!\!$ {\em {Instituci\'o Catalana de Recerca i Estudis  
Avan\c{c}ats (ICREA) and\\ Institut de F\'isica d'Altes Energies, Universitat Aut{\`o}noma de Barcelona\\
08193 Bellaterra, Barcelona, Spain}}

\end{center}

\vspace{0.8cm}

\centerline{\bf Abstract}
\vspace{2 mm}
\begin{quote}\small
We compute the graviton Kaluza-Klein spectrum on a gravity-dilaton background with a naked singularity for all possible boundary conditions at the singularity which are consistent with unitary evolution. We apply methods from non-relativistic quantum mechanics with singular Schr\"{o}dinger potentials. In general the spectrum contains a tachyon, a sign of instability. Only for a particular boundary condition at the singularity is the spectrum free of  tachyons. In this case the lowest-lying graviton mode is massless. We argue  that this result will also hold for other  backgrounds with similar geometry near the curvature singularity. We complete our study with a brief discussion on radion perturbations and the Higgs mechanism on this singular background.    

 \end{quote}

\vfill

 \newpage

\section{Introduction}
Models with a warped extra dimension, such  as those introduced by Randall
and Sundrum~\cite{Randall:1999ee}, offer a geometric solution to the
hierarchy problem. This setup (RS1) originally consisted of ultraviolet (UV) and infrared (IR) branes both embedded in a five-dimensional (5D) space with Standard Model fields   
localized on the IR brane.  The main feature of RS1 is that the electroweak scale
gets suppressed as the volume element on the IR brane becomes exponentially small.
 
With the development of the AdS/CFT correspondence~\cite{Maldacena:1997re,Gubser:1998bc},
it was realized that the strong coupling limit of gauge theories can be described using  a  perturbative \hyphenation{ higher-di-mension-al}
 higher-dimensional gravitational model. In this context, the Randall-Sundrum  geometry is associated with the ${\cal N}=4$ SYM theory. The holographic dual of QCD was constructed~\cite{Erlich:2005qh} following the inverse argument, an approach 
widely  known as AdS/QCD. The conformal symmetry of the $AdS_5$ geometry
corresponds to the conformal limit of QCD at high energies. The  usual IR brane is introduced  in order to break conformal symmetry at low energies. 

An interesting modification of standard AdS/QCD is to deviate from 
the $AdS_5$ metric using a (dilaton) scalar field. Conformal symmetry is
required at high energies. Thus the metric is chosen to
be asymptotically $AdS_5$ near the UV boundary. On the other hand the geometry  
becomes significantly different from $AdS_5$ near the IR. In these
\emph{soft-wall} models conformal symmetry  is gradually broken
giving a more elaborate description of strong 
coupling dynamics and confinement~\cite{Csaki:2006ji}.  
This situation is reminiscent of the Goldberger-Wise mechanism~\cite{Goldberger:1999uk}, where a scalar field is used for the stabilization of the position of the IR brane.

Solutions of gravity coupled to the dilaton typically have a naked singularity at a finite distance  from the UV brane. The introduction of the IR brane can thus be avoided in
this case as the geometry ends naturally at the position of the singularity.  These singular models were initially introduced in an effort to explain the smallness of the cosmological constant~\cite{Kachru:2000hf}. It was subsequently understood that this singularity had to be resolved by a yet unknown stringy configuration, possibly equivalent to a 3-brane, in a way that spoils
the self-tuning of the cosmological constant~\cite{Forste:2000ps}.
One reason for using \emph{soft-wall} models is that, in contrast to
standard AdS/QCD,  they can predict a linear Regge spectrum for the masses
of hadrons and glueballs~\cite{Karch:2006pv,Gursoy:2007er,Batell:2008zm}. This property is intimately related to the behaviour of the warp factor near the singularity. 

The application of \emph{soft-wall} models is not only restricted
to QCD. The hierarchical flavour problem in extra-dimensional models can
be addressed by allowing Standard Model fields to propagate in the
bulk \cite{bulkfields}. In this context it is also possible to describe
electroweak symmetry breaking by a strongly coupled sector using AdS/CFT
correspondence (for reviews see~\cite{Contino:2010rs,Gherghetta:2010cj}).
Gauge bosons propagating in five dimensions acquire a Kaluza-Klein (KK) spectrum
but  the mass of the first excited KK mode is constrained by
electroweak precision tests to be very high~\cite{Huber:2000fh}.  The use of
\emph{soft-wall} backgrounds  made it possible to relax this
stringent constraint allowing the mass of KK excitations to get
as low as $\mathcal O$(TeV)~\cite{Falkowski:2008fz,Batell:2008me,Cabrer:2011fb,Carmona:2011ib,deBlas:2012qf}. 

Some difficulties in \emph{soft-wall} models arise because of the naked singularity. Imposing boundary conditions at the singularity is not straightforward. This situation is similar to non-relativistic quantum mechanics when the Schr\"{o}dinger potential is singular,
e.g.~for a Coulomb potential~\cite{singular}. The usual approach in this case is to demand \textit{boundary conditions that preserve unitarity}. Since wave functions, or their derivatives, generally diverge at the singularity it is impossible to impose boundary conditions in the usual Robin form  ($\phi'=a\phi$), which guarantees unitarity. Alternatively it is possible to fix the  
ratio of the linear combination of solutions ($c_1/c_2$) of the time-independent Schr\"{o}dinger-like equation that describes the various  modes~\cite{Case,barry}. This procedure is also used in~\cite{Wald:1980jn,Brax:2001cx,Horowitz:1995gi} to study waves on singular gravitational backgrounds.  

In this article we will reanalyze the gravitational and Higgs spectra in the specific \emph{soft-wall} model described in Ref.~\cite{Cabrer:2009we} by imposing unitarity on the bulk solutions. In section~\ref{background} we review the gravitational and dilaton background we will be using in the rest of the paper. In section~\ref{graviperturbations} we study the gravitational spectrum for \emph{all possible} boundary conditions at the position of the singularity which are \emph{consistent with unitarity}. In order to do this we use a convenient method to include all possible KK spectra in the same plot. Using this tool we see that a typical KK spectrum contains a tachyon. This is possible as, contrary to the RS1 model, the Schr\"{o}dinger operator
for gravitons is not positive definite in this case. In fact, we find that only
for one specific boundary condition are there no tachyons, in which case the KK 
spectrum contains a massless mode! This is an attractive feature since we can explain 
the tuning required in order to obtain a massless graviton. The essence of the argument is that every other choice is excluded because of the existence of a tachyon which renders the geometrical background unstable. 
         The gravitational background we study depends on a free parameter $\nu$. Depending on $\nu$ it is possible to have  zero, one, or two independent  normalizable eigenfunctions of the 
Schr\"{o}dinger operator that describes  gravitons. The usual approach is to discard non-normalizable modes from the spectrum. This is equivalent to specific boundary conditions.  In Ref.~\cite{singular} it is explained how the existence of non-normalizable solutions is a signal that the actual KK spectrum depends on the details of the resolution of the singularity.  Given such  a resolution the originally non-normalizable modes will be smoothed out. From this point of view constructing the spectrum using only normalizable solutions is not mandatory, but rather a choice not to include modes that depend on the details of the resolution. Of course the most straightforward way to regularize the non-normalizable solutions  is introducing a second IR brane at a small but finite distance from the position of the singularity~\cite{Cabrer:2011fb}. In this article we mainly focus on the case where both solutions are normalizable, with the exception of radion perturbations, where  only one solution is normalizable for all the values of $\nu$ that we consider.   
In section~\ref{sec:Higgs} we give  a brief description of the Higgs mechanism on the singular \emph{soft-wall} background. It is shown that Higgs fluctuations can be treated in the same way as gravitons. In this case it is possible to have a KK spectrum that is free of tachyons and which has a non-zero lowest mass. Finally section~\ref{conclusion} is devoted to our conclusions and outlook. A somewhat technical review of some facts concerning unbounded differential operators and their spectrum is given in appendix~\ref{appendix}.

\section{The Gravitational Background}
\label{background}
We consider the 5D action of gravity, a 3-brane and a dilaton field. The tension of
the 3-brane depends on the value of the dilaton field $\phi$. In the Einstein frame we have   
\be
S=M^3\int d^5x \sqrt{-g} \left(R-\frac{1}{2}\partial_M \phi \partial^M \phi-V(\phi) \right)
-M^3 \int d^4x \sqrt{- \bar g}\, \lambda (\phi),
\label{action}
\ee
where  $\bar g$ is the determinant of the induced metric on the 3-brane and $M$ is the 5D Planck scale. The dilaton field $\phi$  in the above action is dimensionless. We assume that the metric is of the form
\be
ds^2=e^{-2A(y)}\eta_{\mu \nu}dx^{\mu}dx^{\nu}+dy^2,
\label{metric}
\ee
with $\eta_{MN}=\diag\{-1,1,1,1,1\}$. The brane is  located at the  $y=0$ hypersurface and we impose  a $\mathbb{Z}_2$ orbifold symmetry $y \to -y $.  A variation of the action with respect to $g_{M N}$ gives
\be
G_{M N}=\frac{1}{2}T_{M N}-
\frac{1}{2} \frac{\sqrt{-\bar g}}{\sqrt{-g}} \bar g_{\mu \nu} \delta_M^{\mu} \delta_N^{\nu}
\lambda (\phi) \delta (y). 
\label{ein1} 
\ee
The  equations of motion for the ansatz of Eq.~(\ref{metric})  are
\bea
3A''(y)&=&\frac{\phi'(y)^2}{2}+ \frac{\lambda (\phi) \delta (y)}{2}, \label{eqm1}  \\
6 A'(y)^2&=&-\frac{V(\phi)}{2}+\frac{\phi'(y)^2}{4}  \label{eqm2},\\
\phi''(y)-4A'(y)\phi'(y)&=&V'(\phi)+\lambda'(\phi)\, \delta(y) \label{eqm3}. \\ \nn
\eea
where $f^\prime$ denotes derivative with respect to the function argument. 
Boundary conditions on the UV brane are calculated by integrating in
a small interval around $y=0$,
\bea
\left. A' \right|_{0^+}-\left. A' \right|_{0^-}&=&\frac{\lambda[\phi(0)]}{6},  \\ 
\left. \phi' \right|_{0^+}-\left. \phi' \right|_{0^-}&=&\lambda'[\phi(0)].  
\eea
Taking into account the  $\mathbb{Z}_2$ symmetry around the brane  we have
\bea
\left. A' \right|_{0^+}&=&\frac{\lambda[\phi(0)]}{12},  \label{bou1} \\ 
\left. \phi' \right|_{0^+}&=&\frac{\lambda'[\phi(0)]}{2}.  \label{bou2}
\eea

The system of Eqs.~(\ref{eqm1})-(\ref{eqm3}) with the above boundary conditions
has a unique solution, given the potentials $\lambda(\phi)$, $V(\phi)$ 
and the value of $\phi(0)$~\footnote{We can choose  $A(0)=0$ without  loss of generality.}.
Following~\cite{DeWolfe:1999cp}  we introduce the ``superpotential" $\su(\phi)$, defined by 
\be
V=18 \left( \frac{\partial \su}{\partial \phi} \right)^2-12 \su^2.
\ee
Background  solutions can be generated  using $\su$ as 
\bea 
A'(y)&=&\su(\phi),  \label{solW} \\
\phi'(y) &=& 6 \su^{\,\prime}(\phi),
\eea
while the boundary conditions are satisfied if
\bea 
\su(\phi(0))&=&\frac{\lambda[\phi(0)]}{12}  \label{bcW} \\
\su^{\,\prime}(\phi(0)) &=&  \frac{\lambda'[\phi(0)]}{12},
\eea

Choosing $\su$ is not completely arbitrary. The asymptotic form of the metric near
the UV brane must approach AdS$_5$. Additional pathologies appear 
when the scalar potential $V(\phi)$ is not bounded above for the background 
solution~\cite{Gubser:2000nd}~\footnote{This is known  in the literature as the Gubser criterion.}.

Considering the above restrictions, we will consider the superpotential in Ref.~\cite{Cabrer:2009we} which reads as
\be 
\su=k \left(1+e^{\nu \phi/\sqrt{6}} \right).
\label{superpot}
\ee
The resulting  background has an asymptotic form near the singularity which has been often considered in the literature~\cite{Karch:2006pv,Gursoy:2007er,Batell:2008zm}. The background 
generated in this case reads as  
\bea
A(y)&=&ky-\frac{1}{\nu^2}\log \left(1-\frac{y}{y_s} \right), \label{backsol1} \\
\phi(y)&=&-\frac{\sqrt{6}}{\nu} \log \left[k \nu^2 \left(y_s-y \right) \right], \label{backsol2} 
\eea
for $0<y<y_s$. The above expressions also hold in the $-y_s<y<0$ interval,  with the
replacement  $y \to -y$, due to the $\mathbb Z_2$ symmetry. There is a naked  curvature singularity  at a finite coordinate distance $y_s$ from the UV brane. This is to be considered as the end of spacetime. The dilaton also blows up at $y_s$. The location of the singularity $y_s$ depends exponentially on the brane value of $\phi$ as $ky_s=\frac{1}{\nu^2}\exp[-\nu\phi(0)/\sqrt{6}]$. All we need to create the weak hierarchy is $\phi(0)<0$ and otherwise $|\phi(0)|=\mathcal O(1)$ which can be achieved with a fairly generic brane potential~\cite{Cabrer:2009we}.

The low energy  4D Lagrangian is calculated by integrating the action along the fifth
dimension for the background configuration. This is equivalent to  an effective 4D
cosmological constant,       
\be
\Lambda_{eff}=\frac{2}{3}\int_{-y_s^+}^{y_s^-} dy e^{-4A(y)}V(\phi)+\frac{1}{3}e^{-4A(0)}\lambda[\phi(0)],
\ee
where we have used the equations of motion (\ref{eqm1})-(\ref{eqm3})
in the scalar kinetic and curvature terms of the action and we are omitting the global $M^3$ factor. The above formula can be simplified by using  the identity 
\be
\int dy e^{-4A(y)}V(\phi)=3 \int \frac{d}{dy} \left(  e^{-4A(y)} \su[\phi(y)] \right),
\ee
and condition (\ref{bcW}). The final  result is 
\be
\Lambda_{eff}=4 e^{-4A(y_s^-)} \su [\phi(y_s^-)].
\ee

Substituting  $A(y)$ and $\phi(y)$ from Eq.~(\ref{backsol1}) we see that $\Lambda_{eff}=0$ for $\nu<2$. In this case the background is consistent without a contribution to the 4D cosmological constant from the singularity. For $\nu=2$ we have 
\be
 \Lambda_{eff}= k\,\frac{e^{-4 k y_s}}{4ky_s}.
\ee
This is in conflict with the metric ansatz of Eq.~(\ref{metric}) which
is flat from the 4D perspective. Consistency requires a contribution to
the cosmological constant from the position of the singularity that would
exactly cancel $\Lambda_{eff}$. This corresponds to the usual tuning 
of the cosmological constant in extra-dimensional models~\cite{Forste:2000ps}.
It is interesting to notice that when the Gubser criterion gets violated for $\nu>2$, 
$\Lambda_{eff}$ is infinite. It is hard to imagine how such a contribution could be 
canceled in order to have a consistent background~\cite{Kim:2000yq}. For this reason
in this article we will consider backgrounds with $\nu \leq 2$~\footnote{Of course the 
$\nu>2$ case can be made consistent by assuming that the singularity is resolved by introducing  a second brane at $y_1=y_s-\epsilon$, with $\epsilon>0$ a small but finite coordinate distance from $y_s$~\cite{Cabrer:2010si}. }.

\section{Gravitational Perturbations}
\label{graviperturbations}

In the previous section we have introduced two scalars, the metric $A$ and the stabilizing scalar field $\phi$, whose background values define the geometry of the 5D space time, as well as the constant graviton background $\eta_{\mu\nu}$. In this section we will study fluctuation of the quantum fields around the previous background values which give rise to the respective KK modes which define the 4D spectrum of the gravitational sector.

\subsection{Graviton spectrum} 
In this section we proceed with the calculation of the KK spectrum of gravitational fluctuations
in the background given by  Eqs.~(\ref{backsol1})-(\ref{backsol2}) and whose massive modes can be interpreted as the spectrum of excited composite states of the strongly coupled dual field  theory.  Metric perturbations can be decomposed, according to their transformation properties, to a transverse traceless tensor, a four-vector and two scalars. In this section we focus on the tensor (graviton) part of the spectrum. Transverse traceless graviton perturbations are defined  by
\be
ds^2=e^{-2A}\left(\eta_{\mu \nu}+h_{\mu \nu} \right)dx^{\mu}dx^{\nu}+dy^2,
\ee
with $\partial^{\mu}h_{\mu \nu}=0$ and $h^{\mu}_{\, \mu}=0$. A separation of variables is performed  assuming solutions of the form $h_{\mu \nu}(x^{\mu},y)=h_{\mu \nu}(x^{\mu})h(y)$ where the sum over KK modes is left implicit.

Following~\cite{Csaki:2000zn,Kiritsis:2006ua} the KK spectrum of  gravitons is described by the eigenvalue problem  
\be
\hat{\mathcal O}\, h(y)=m^2 h(y).
\label{sturm}
\ee 
where we have defined the operator
\be
\hat{\mathcal{O}} \equiv-e^{-2A} \frac{d^2}{dy^2}+4e^{-2A} A'\frac{d}{dy},
\label{operator}
\ee
which can be self-adjoint on a Hilbert space with norm 
\be
(f,g)=\int e^{-2A} f^* g\, dy.
\label{norm}
\ee 
Defining $\tilde{h}=he^{-2A}$ we can write Eq.~(\ref{sturm}) as
\be
 -\tilde{h}''+V(y)\tilde{h}=m^2e^{2A} \tilde{h} ,
\label{schro}
\ee 
where   the  potential $V(y)$ is given by
\be
V(y)=4 A'^2(y)-2A''(y). \label{pot}
\ee 
and for the metric (\ref{backsol1}) it takes the  form 
\be
V(y)=4 k \left(k+\frac{2}{\nu^2 (y_s-y)}\right)-\frac{2 \left(\nu^2-2\right)}{\nu^4 (y-y_s)^2} \label{potfull} \,.
\ee
The criterion of normalizability becomes in this case   
\be 
\int_0^{y_s} e^{2A} |\tilde{h}|^2 dy<\infty.
\ee

It is useful to examine the form of Eq.~(\ref{schro}) dropping subdominant terms
near the singularity as
\be \tilde{h}''(y)+
\frac{2
   \left(\nu^2-2\right) \tilde{h}(y)}{\nu^4
   (y-y_s)^2}+m^2 \tilde{h}(y)
   \left(1-\frac{y}{y_s}\right)^{-\frac{2}{\nu^2}}=0.
\label{schroasy}
\ee
When $\nu\leq 1$  the dominant terms   of Eq.~(\ref{schroasy}) near
the singularity are the first and the third. Both linearly independent
solutions  are non-normalizable near the singularity. This is in agreement with Ref.~\cite{Cabrer:2009we}. There it is shown that, working in conformal coordinates defined as 
\be 
dz=e^{A(y)}dy
\ee
and with $\nu<1$,  the position of the singularity moves to infinity. The 
Schr\"odinger  potential for the graviton in this frame is almost zero for sufficiently large $z$. The eigenfunctions behave asymptotically like plane waves. These are non-normalizable functions  in an infinite interval that therefore do not belong to the physical spectrum. On the other hand,  it is possible to build  normalizable functions as linear combinations of plane waves because they form a complete basis in function space.  In complete analogy with the spreading of the Gaussian wave packet in quantum mechanics, their evolution is unitary but the probability distribution spreads with time. Since there is no discrete spectrum of mass eigenfunctions, fluctuations in the  $\nu<1$ case can be interpreted as unparticles.  

When  $\nu>1$  the dominant balance near the singularity is between the first and the second term. The general solution can be written in general as a linear  combination $\tilde{h}(y)=c_1 g_1(y)+c_2 g_2(y)$ of two  linearly independent solutions $g_1,g_2$, with the behaviour  
\be
\tilde{h} \sim c_1\left[k(y_s-y)\right]^{1-2/\nu^2}+c_2\left[k(y_s-y)\right]^{2/\nu^2},
\label{asymp}
\ee
near the singularity. For $1<\nu<\sqrt{2}$ only the second term of (\ref{asymp}) is normalizable. The spectrum of $\Op$ can be restricted to contain only functions with $c_1=0$. This condition is equivalent to the requirement that $e^{-4A(y_s)} h'(y_s)=0$ and the lowest KK mode is massless.  One can notice that the Schr\"{o}dinger potential $V(y)$ is in this case repulsive and it behaves
like $V(y) \sim c/(y_s-y)^2$ with $0<c<2$ near  $y=y_s$.

It is important to realize that the resolution of the singularity would regularize the singular Schr\"{o}dinger potential near $y=y_s$ normalizing the previously  non-normalizable solution. It is then possible to argue that the choice of $c_1/c_2$  is actually arbitrary even for $1<\nu<\sqrt{2}$, but only for $c_1=0$ can we  compute the mass spectrum without knowing the details of the 
singularity resolution. Notice that eigenfunctions that contain the non-normalizable term will behave like $(y_s-y)^{1-2/\nu^2}$ near the singularity. Let us now assume that we wish to normalize such eigenfunctions e.g.~by introducing a normalizing factor $N$.
Such a normalizing factor would behave like $N \sim \ell^{\,(3/\nu^2-3/2)}$, where $\ell$ is the cutoff scale of  the resolution. For $\nu<\sqrt{2}$ the normalizing factor $N$ goes to zero when $\ell$ goes to zero. As a result non-normalizable eigenfunctions would correspond to states that are highly localized towards the singularity when the resolution is introduced. If this resolution is artificial then non-normalizable states would also be artifacts of the cutoff.

Things become different when $\nu>\sqrt{2}$. Both linearly independent solutions
of Eq.~(\ref{schro}) are normalizable. The $y$ derivative of the $c_1$ term is divergent at $y_s$, so it is impossible to impose boundary conditions using the Robin form. What can be done 
is to choose a constant value for the ratio $c_1/c_2$.  This condition is consistent
with unitarity and will provide us with a discrete mass spectrum for $\Op$. The proof of 
this statement can be found in the Appendix. There are two independent ways to calculate the KK spectrum when $\nu>\sqrt{2}$.  

\subsubsection{Method I}

\begin{figure}[t]
\hspace{-1cm}
\begin{tabular}{cc}
\epsfig{file=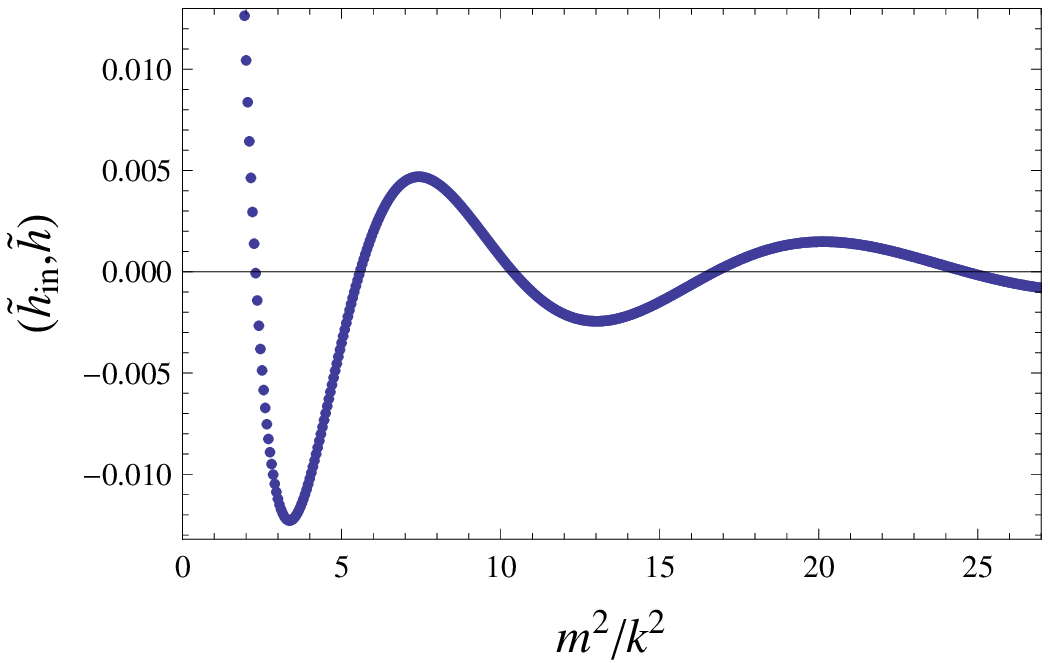,width=0.525\linewidth,clip=} & 
\epsfig{file=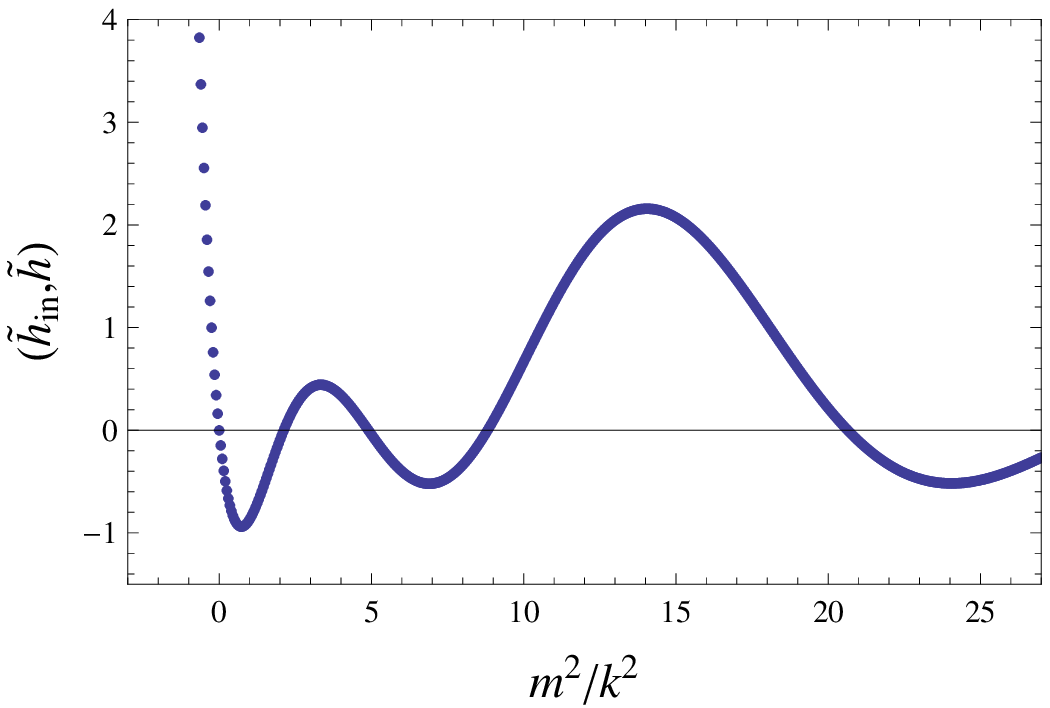,width=0.50\linewidth,clip=} \\
\end{tabular}
\caption{\it On the left panel we plot $ (\tilde{h}_{in}, \tilde{h})$ as a function of $m^2/k^2$ with
initial conditions $h(0)=1,h'(0)=0$. We have chosen $ky_s=1$ and $\nu=1.5$.  The eigenfunction $\tilde{h}_{in}$ corresponds to $m_{in}^2=0$. The  zeroes $m^2_i$ of this plot correspond to  the spectrum of  mass eigenvalues. On the right panel we plot the same but for $m_{in}^2=14k^2$. 
}
 \label{figortho}
 \end{figure}
 
The first method uses the fact that  self-adjoint operators must have an orthonormal set of eigenfunctions which form a complete basis. This  can be used in order to build an algorithm
to compute the mass spectrum. Initially an arbitrary value for $m^2$ is chosen, let us say $m^2_{in}$. The Neumann condition $h'(0)=0$, or the equivalent one
 \be
\tilde{h}'(0)=-2 \left(k+\frac{1}{\nu^2y_s} \right) \tilde{h}(0)
\label{boundary}
\ee
on the regular brane, along with $\tilde{h}(0)=1$, is sufficient to uniquely determine a solution $\tilde{h}_{in}$ of Eq.~(\ref{schro}). As in most cases an analytical form is usually difficult to obtain, so it is convenient to use a numerical algorithm. Solution $\tilde{h}_{in}$ corresponds to a definite value for $c_1/c_2$ or equivalently to a boundary condition at the singularity. All the modes in the spectrum that correspond to $m^2_{in}$ must then be consistent with this boundary condition. 

The value of the mass is then increased (and/or decreased) until we find an $m^2=m^2_n$ with a corresponding eigenfunction $\tilde{h}_{n}$ that satisfies
\be
(\tilde{h}_{in},\tilde{h}_n)=\int e^{2A(y)} \tilde{h}_{in}^* \tilde{h}_{n} dy=0.
\ee
This is a new  eigenfunction with eigenvalue $m_n^2$. Repeating this procedure for higher (and/or lower) values of the mass,  the whole spectrum of $\Op$ can be calculated. There is no need to estimate the ratio $c_1/c_2$ at any point as the self-adjointness of the operator is established  through the orthogonality of the eigenfunctions. As an example in Fig.~\ref{figortho} we can see the spectrum for $ky_s=1$ and $\nu=1.5$ assuming an initial mass $m_{in}^2=0$ (left panel) and $m_{in}^2=14 k^2$ (right panel). The mass eigenvalues correspond to the zeroes of $ (\tilde{h}_{in}, \tilde{h})$.

Throughout this article the values of $ky_s$ for the numerical calculations have been chosen arbitrarily since we did not aim to perform any phenomenological analysis. As far as we can tell, the main conclusions of this article hold for every $k \geq0$ and $y_s>0$. Further discussion on this can be found  below.

\begin{figure}[t]
\hspace{-1cm}
\begin{tabular}{cc}
 \epsfig{file=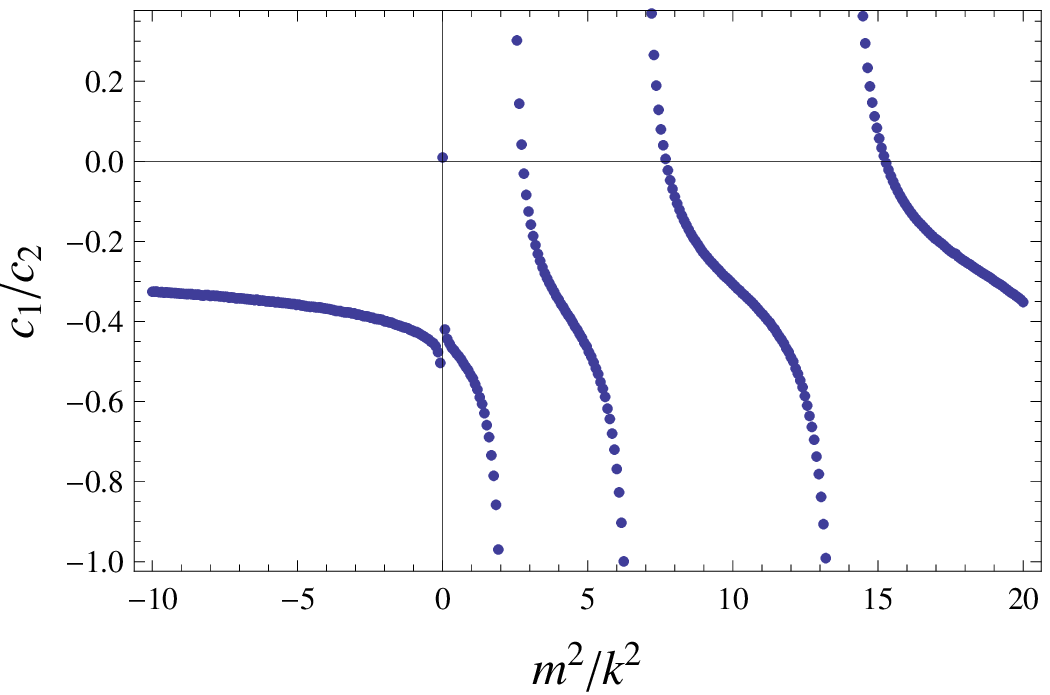,width=0.50\linewidth,clip=}& 
\epsfig{file=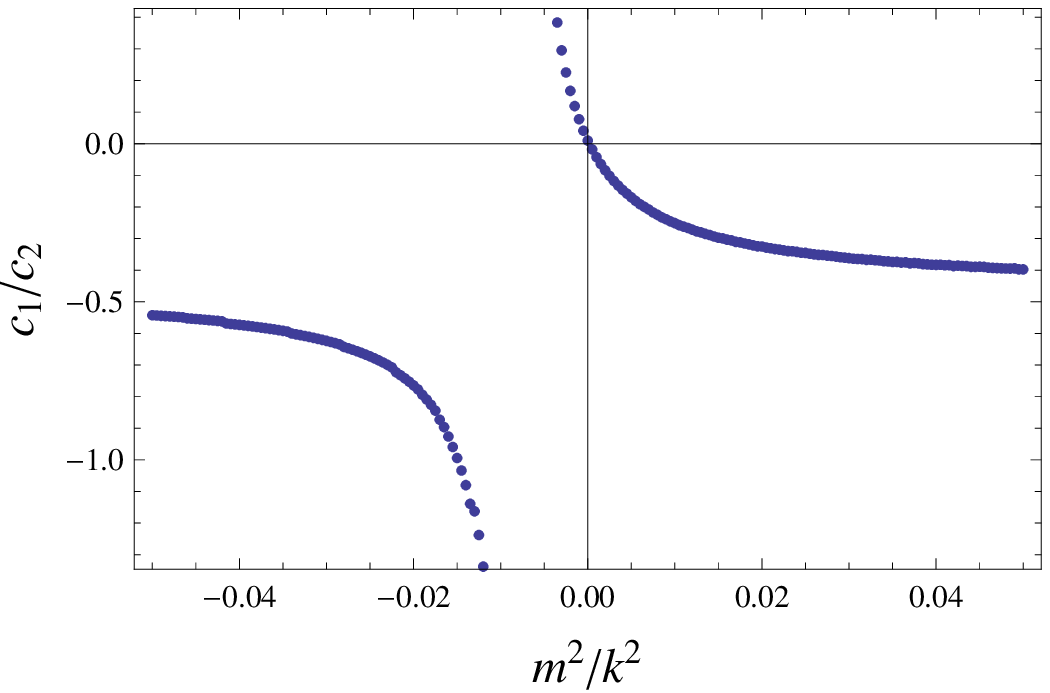,width=0.50\linewidth,clip=}\\
\end{tabular}
\caption{ \it Left panel:  plot of the value of $c_1/c_2$ as function of $m^2/k^2$ for $ky_s=1$, $\nu=1.8$. 
Right panel:  the same but  focused on the $m^2 \sim 0$ region. }
 \label{c1c2}
 \end{figure}

\subsubsection{Method II}
A second method to calculate the KK spectrum is to solve Eq.~(\ref{schro})  numerically for a range of values of $m^2$ with the same initial conditions  $h(0)=1$ and $h'(0)=0$ at $y=0$  and then    calculate the ratio $c_1/c_2$ from the asymptotic behaviour of these solutions. In order to do
this we focus on a small interval very close to $y_s$ and perform a fit of our solution using  Eq.~(\ref{asymp}). The result of this procedure is shown in the left panel of Fig.~\ref{c1c2}, where we see $c_1/c_2$ as a function of $m^2/k^2$. In order to  compute the spectrum, for fixed $c_1/c_2\equiv c_{12}$, we have to \emph{draw a horizontal line} on the left panel of Fig.~\ref{c1c2}
and compute the solution of the equation $c_1/c_2[m^2]=c_{12}$. The $m^2 \sim 0$ region of Fig.~\ref{c1c2} is not clearly visible. In the right panel of Fig.~\ref{c1c2} we zoom it in order to demonstrate that the function $c_1/c_2[m^2]$ crosses the origin. Thus for $c_{12}=0$  the spectrum contains a massless mode.  

\begin{figure}[htb]
\centering
 \epsfig{file=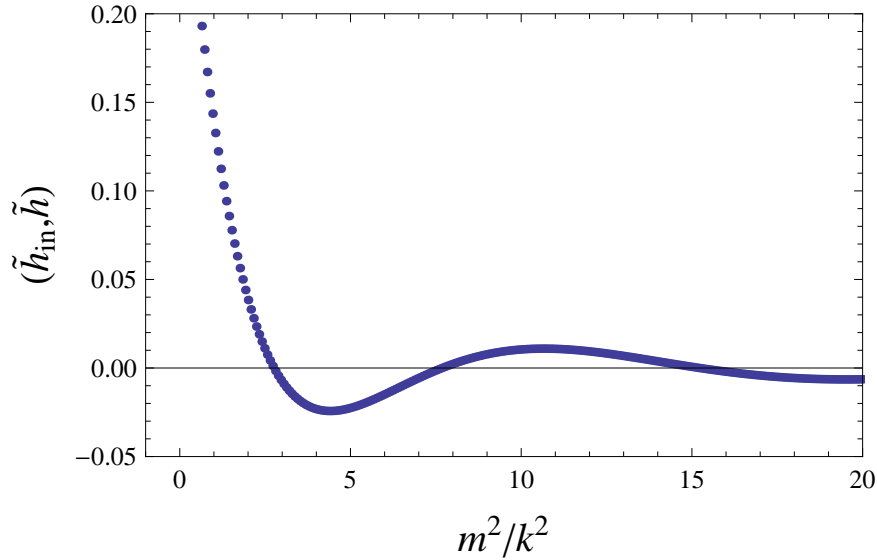,width=0.7\linewidth,clip=}
\caption{ \it Plot of $(\tilde{h}_{in}, \tilde{h})$ for $m_{in}^2=0$, $ky_s=1$ and $\nu=1.8$.}
\label{ds}
 \end{figure}
The validity of this numerical procedure can be confirmed by a cross-check using the first method we discussed. For example in Fig.~\ref{ds} we plot $(\tilde{h}_1, \tilde{h})$ for $m^2_{in}=0$, which corresponds to $c_{12}=0$. The zeroes of Fig.~\ref{ds} match those of Fig.~\ref{c1c2} implying that the spectra coincide. An advantage of studying Fig.~\ref{c1c2} is that {\it one can inspect every possible spectrum of $\Op$} for every consistent choice of boundary conditions on the singularity.

The structure of the $c_1/c_2$ curves around $m^2=0$ is of particular interest. In the right panel of Fig.~\ref{c1c2} we have zoomed in the $m^2\sim0$ part of its left panel in order to reveal some hidden features.  There is a branch that crosses the origin and has negative $m^2$ with $c_{12}>0$. The negative $m^2$ modes correspond to tachyons which render  the particular  background unstable. A second branch with $ m^2 \lesssim -0.01k^2$ also exists. If $c_{12} \to 0$ as $m^2 \to -\infty$ then the spectrum is tachyon-free only for $c_{12} \to 0$. In this case the lowest mass is $m^2=0$.

\begin{figure}[htb]
\begin{center}
 \epsfig{file=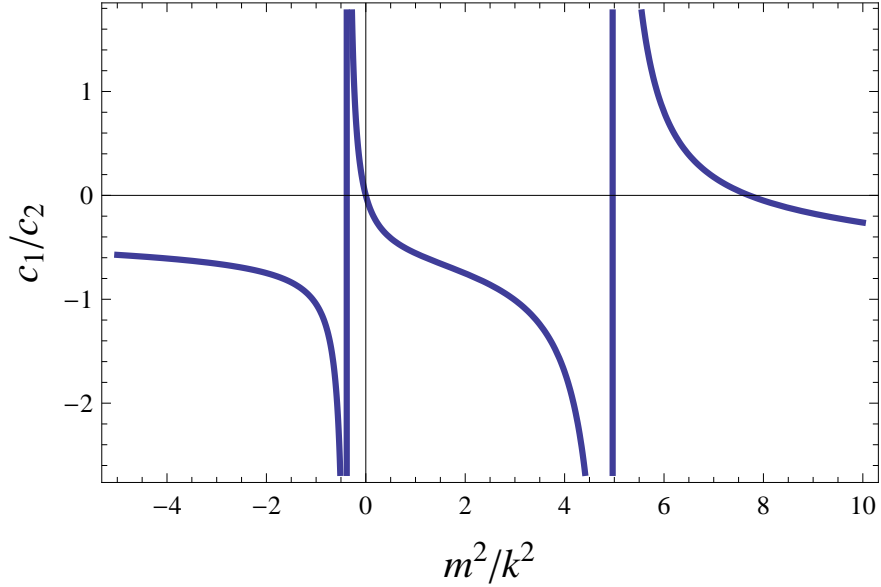,width=0.7\linewidth,clip=}
\end{center}
\caption{$c_1/c_2$ as function of $m^2/k^2$ calculated analytically for the simplified metric (\ref{simpmetric}), $ky_s=1$ and $\nu=\sqrt{3}$. Horizontal lines mark the position of poles.}
 \label{analytical}
 \end{figure}

Performing numerical fits, we  indeed verify that the  ratio  $c_1/c_2$  behaves as $(-k^2/m^2)^{1/4}$ for $m^2/k^2 \to -\infty$. As the numerical approach could be debatable, since numerics 
become untrustworthy for  large values of $|m^2|$, we will confirm this conclusion in a case where there exists an analytical solution. To this end we will consider the simplified metric
\be
A(y)=-\frac{1}{\nu^2}\log\left( 1-\frac{y}{y_s}\right)
\label{simpmetric}
\ee
stemming from the superpotential 
\be
\mathcal W=ke^{\nu\phi/\sqrt{6}}
\label{simpsuper}
\ee
The metric (\ref{simpmetric}) retains the main features of the complete metric (\ref{backsol1}) near the singularity, although it departs from it near the UV brane and thus it would not be a good candidate to explain the weak/Planck hierarchy. The potential in Eq.~(\ref{schro}) corresponding to the metric in Eq.~(\ref{simpmetric}) is given by
\be
V(y)=-\frac{2 \left(\nu^2-2\right)}{\nu^4 (y-y_s)^2} \label{simppot} \,.
\ee
which formally corresponds to keeping only the last term of the full potential (\ref{potfull}) or the zeroth order in the expansion in powers of $k$. Now Eq.~(\ref{schro}) for $\nu=\sqrt{3}$ can be solved analytically. The result  is a  linear combination of Bessel functions as
\be
h=c_1\frac{  \sqrt{x} \, I_{-\frac{1}{4}}\left(\frac{3
   x^{2/3}}{2 \sqrt{\epsilon }}\right)}{ \epsilon
   ^{3/8}}+c_2\frac{  \sqrt{x} \, I_{\frac{1}{4}}\left(\frac{3
   x^{2/3}}{2 \sqrt{\epsilon }}\right)}{\epsilon^{3/8}},
\ee
where $x= k^{3/2}(y_s-y)y_s^{1/2}$ and $\epsilon=-k^2/m^2$. Applying the usual  conditions on the regular brane and after a Taylor expansion of the solutions for $\epsilon \to 0$, we confirm that the leading behaviour is $c_1/c_2 \sim \epsilon^{1/4}$. Moreover in Fig.~\ref{analytical} $c_1/c_2$ is calculated analytically for $\nu=\sqrt{3}$ as a function of the mass. One can easily check that the branch structure in Fig.~\ref{analytical} is similar to the one in the left panel of  Fig.~\ref{c1c2}. 

We have repeated the numerical calculation of the $c_1/c_2$ plot for various values of $0 \leq  ky_s \leq 8$, where our numerical methods are accurate, and $ \sqrt{2}<\nu<2$ getting the same result: only for $c_1/c_2=0$ the spectrum is tachyon free. This is not a surprise as stability is related to the low-lying eigenvalues of Eq.~(\ref{schro}), and the low-lying modes  are dominated by the most negative term of $V(y)$, which is the last, $k$-independent, term of Eq.~(\ref{potfull}) near the singularity. The existence of tachyons relies on the asymptotic form of the potential and the asymptotic behaviour of the solutions near the singularity, which is given by the ratio $c_1/c_2$. The $k$-dependent term in the potential $V(y)$ is only important for higher KK modes. An additional $k$-independent  feature is that the $c_1/c_2$ curves cross the origin. For $m^2=0$ and $h'(0)=0$ the solution of Eq.~(\ref{sturm}) is $h(y)=constant$, which leads  to $c_1=0$. Taking into account the accumulated evidence, we believe that it is safe to conclude that the graviton KK spectrum contains a tachyon for $c_1 \neq 0 $ and any value of $ky_s$. 

\subsection{Scalar perturbations}
The analysis of perturbations on the \emph{soft-wall} background is completed with the study of scalar radion and dilaton perturbations. Vector perturbations can be gauged away except for a possible zero mode~\cite{Kiritsis:2006ua}.  Following the study  of  Ref.~\cite{Csaki:2000zn}  (see also Ref.~\cite{Cabrer:2009we})  scalar perturbations are defined by
\bea
\phi(x,y)\,&=&\phi(y)+\varphi(x,y), \label{dilpert} \\
ds^2&=&e^{-2A(y)-2F(x,y)} n_{\mu \nu}dx^{\mu}dx^{\nu}+(1+G(x,y)^2)dy^2, \label{scalper}
\eea
where  $\varphi(x,y)$ is the dilaton perturbation and $F(x,y)$ and $G(x,y)$ are the scalar gravitational perturbations. Not all of the above quantities are dynamically independent. The equation of motion of $F$ is given, after a separation of variables,  by
\be
F''-2A'F'-4A''F-2 \frac{\phi''}{\phi'}F'+4A'\frac{\phi''}{\phi'}F=-m^2e^{2A}F. \label{radion}
\ee
Two constraints fix the nondynamical quantities $\varphi$ and $G$   
\bea
\phi' \varphi &=& 6(F'-2A'F), \\
G&=&2F. 
\eea
In order to define a self-adjoint operator from Eq.~(\ref{radion}) it is necessary to use the 
following inner product 
\be
(F_1,F_2)=\int_0^{y_s} e^{-A-\log\phi'}F_1^*\,F_2\,dy.
\ee 
With the redefinition $\tilde{F}=e^{-A-\log\phi'}F$, Eq.~(\ref{radion}) can be 
written in the form
\be
-\tilde{F}''+V_F(y)\tilde{F}=m^2e^{2A}\tilde{F}.
\ee 
Near the singularity the above equation is approximated by
\be
-\tilde{F}''+\frac{1+\nu^2}{\nu^4(y_s-y)^2}\tilde{F}=m^2(y_s-y)^{-2/\nu^2}y_s^{2/\nu^2} \tilde{F}.
\ee
For $1<\nu<2$, the $m^2$ term is subdominant and the  asymptotic form of solutions is  
\be
\tilde{F}=c_1(y_s-y)^{-1/\nu^2}+c_2 (y_s-y)^{1+1/\nu^2}.
\label{asyF}
\ee
Taking into account the normalizability condition for $\tilde{F}$
\be 
\int_0^{y_s} e^{A+log(\phi')} |\tilde{F}|^2 dy<\infty,
\ee
it can be  seen that the $c_1$ term of Eq.~(\ref{asyF}) is non-normalizable for every $\nu$. The spectrum can be computed, without resolving the singular
potential $V_F$, only for $c_1=0$~\cite{Cabrer:2009we}.

The most straightforward way to resolve the singularity is to introduce a brane at $y=y_s-\ell$, where $\ell$ is a small but finite coordinate distance. The resulting spectrum will depend on $\ell$
in addition  to $k$ and $\nu$. A complete understanding of all possible spectra in this case is complicated and beyond the scope of  this article. Nevertheless we present  a typical calculation with an IR brane resolution in order to exhibit how to apply the $c_1/c_2$ plot method in this case.

\begin{figure}[t]
\begin{center}
 \epsfig{file=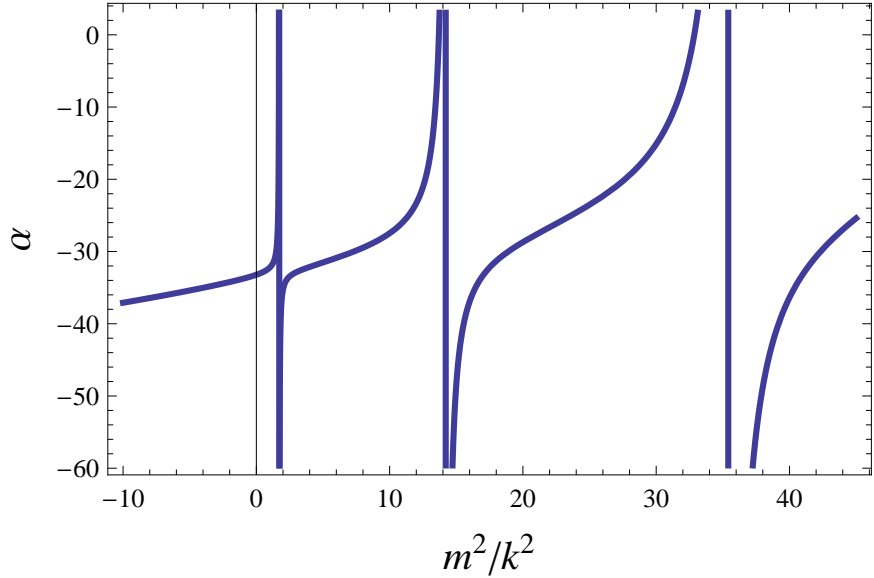,width=0.7\linewidth,clip=}
\end{center}
\caption{ \it Radion KK modes: $\alpha$ as function  of $m^2/k^2$ calculated analytically for the simplified metric of Eq.~(\ref{simpmetric}), $ky_s=1$, $\nu=\sqrt{3}$ and $\ell=10^{-3}$.  }
 \label{radionanalytical}
 \end{figure}

 Indeed it is not necessary to use the $c_1/c_2$ ratio in this case since it is possible to define a boundary condition of the form $\tilde{F}'(y_s-\ell)=\alpha \tilde{F}(y_s-\ell)$ at $y=y_s-\ell$  with  $\alpha \in \mathbb{R}$. Following the same steps as for the $c_1/c_2$ method above, a solution of Eq.~(\ref{radion}) with initial conditions $F(0)=0$, $F'(0)-2A'(0)F(0)=0$  is found for a number of $m^2$ values in an interval. The boundary conditions above correspond to the stiff potential limit ($\lambda''(\phi) \gg 1$). Given a set of  solutions, $\alpha$ is calculated in order to construct an $\alpha=\alpha(m^2)$ plot. The spectrum is given by the intersection points of  this plot with the horizontal line corresponding to a given value of $\alpha$. For example in Fig.~\ref{radionanalytical} we see the $\alpha=\alpha(m^2)$ plot for the simplified metric of Eq.~(\ref{simpmetric}) which yields the potential
 \be
 V_F=\frac{1+\nu^2}{\nu^4(y_s-y)^2}
 \ee
for $ky_s=1$, $\nu=\sqrt{3}$ and $\ell=10^{-3}$. In accordance with the case of gravitons the solution is obtained analytically as a combination of Bessel functions. We see that for $\alpha \lesssim -33$ a tachyon appears in the spectrum rendering the background unstable. 

\section{The Higgs Mechanism}
\label{sec:Higgs}

In this section we present some aspects of EWSB on the \emph{soft-wall} background with a Higgs field propagating in the bulk. The strategy, which is usually followed in the literature in order  to trigger electroweak symmetry breaking (EWSB),  is to introduce a new infrared (IR) brane with a localized Higgs potential. The Higgs then acquires a vacuum expectation value (VEV) which is the coordinate dependent classical solution. Following the method developed in the previous section, it is possible to avoid the introduction of the IR brane. Assuming unitarity,  the ratio $c_1/c_2$ of Higgs fluctuations around a given background is fixed and the KK spectrum of the Higgs field is now well defined.

A 5D Higgs transforming as the $(\textbf{2},1/2)$ representation of $SU(2)_L \times U(1)_Y$ is introduced as
\be
H=e^{i \vec{\chi}(x,y) \vec{\sigma}} 
\left(
\begin{array}{c}
0\\
h(y)+\xi(x,y)\\
\end{array}
\right),
\ee
where the vector $\vec{\chi}$ corresponds to the three 5D Goldstone bosons. The Higgs action contains a gauge invariant kinetic term and a potential $V(\phi,h)$ that couples the
Higgs field with the dilaton. 

The classical Higgs background $h(y)$ can be generated by using the superpotential  formalism. In particular we can assume a superpotential of the form 
\be
\su_{H}=\frac{1}{12}ak \left|H \right|^2,
\ee
where $a$ is a dimensionless constant, which is added to the dilaton superpotential as
\be
\su=\su_{\phi}+\su_{H}=k \left(1+ e^{\nu \phi/\sqrt{6}} \right)  +\frac{1}{12}ak h^2.
\ee
 The superpotential is related to  the Higgs-dilaton potential by
\be
V(\phi,h)=18 \left( \left( \frac{ \partial \su}{\partial \phi} \right)^2+ \left(
\frac{ \partial \su}{\partial h} \right)^2 \right)-12 \su^{\,2}.
\ee
The background solution of the Higgs field is 
\be
h'(y)=6\frac{ \partial \su }{{\partial h}},
\label{higgsbeq}
\ee
which can be easily solved for $h(y)$ as
\be 
h(y)=h_0e^{aky}.
\ee
The dimensionless integration constant $h_0$ can be considered to be small in order  to avoid backreaction to the metric. Nevertheless we will include the backreaction in our description as it could possibly result in interesting  models. The background solution for the dilaton is given in Eq.~(\ref{backsol2}) and the warp factor is then calculated from $A'=\su$ as
\be
A(y)=\frac{1}{24} h_0^2\left(e^{2aky}-1 \right)+ky-\frac{1}{\nu^2}\log\left(1-\frac{y}{y_s} \right).
\ee

The properties of Higgs particles result from the study of fluctuations $\xi(x,y)$, that occur around the classical solution. A separation of variables for $\xi$ is needed in order to compute the KK spectrum: $\xi(x,y)=e^A\mathcal H(x)\xi(y)$ with
\be 
\xi''-2 A' \xi'= \left(\frac{\delta^2 V(\phi,h)}{\delta h^2}-m^2e^{2A}+3A'^2-A''\right) \xi+\frac{\delta^2\lambda(\phi,h)}{\delta h^2}\delta(y) \xi\,,
\ee
where  $m^2$ is the 4D mass eigenvalue.  
It is convenient to define
\be 
m^2_5(y)\equiv\frac{\delta^2 V(\phi,h)}{\delta h^2}=a^2k^2\left(1- h_0^2 e^{2ak y} \right)-4ak-\frac{4ak}{\nu^2 (y_s-y)},
\ee
so that
\be
\xi''-2 A' \xi'= \left(m^2_5(y)-m^2e^{2A}+3A'^2-A''\right) \xi+\frac{\delta^2\lambda(\phi,h)}{\delta h^2}\delta(y) \xi\,,
\label{higgsfluc}
\ee
As in the case of gravitons, the inner product for $\xi$ is given by 
\be
(\xi_1,\xi_2)=\int_0^{y_s}  \xi_1^* \xi_2\, dy \,.
\label{normxi}
\ee

The boundary conditions for the fluctuations $\xi$ are given by integrating (\ref{higgsfluc}) around $y=0$
\be
\frac{\xi'(0)}{\xi(0)}=\left. \frac{\delta^2 \lambda (\phi,h)}{\delta h^2} \right|_{y=0}=2ak
\ee

\begin{figure}[htb]
\hspace{-1cm}
\begin{tabular}{cc}
\epsfig{file=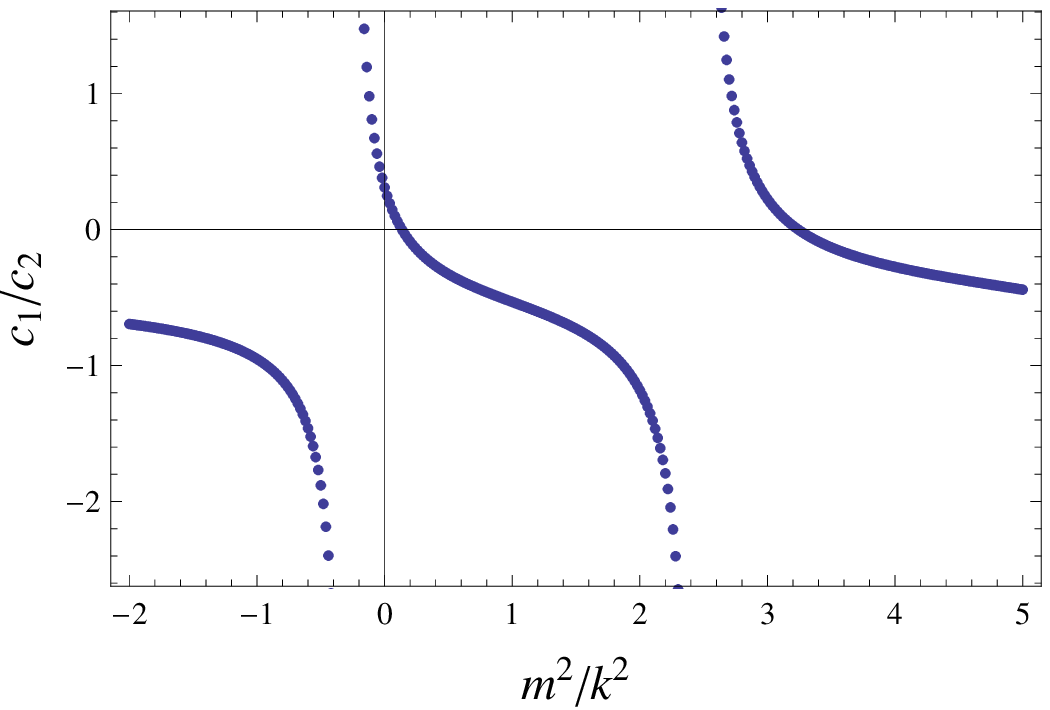,width=0.50\linewidth,clip=} & 
\epsfig{file=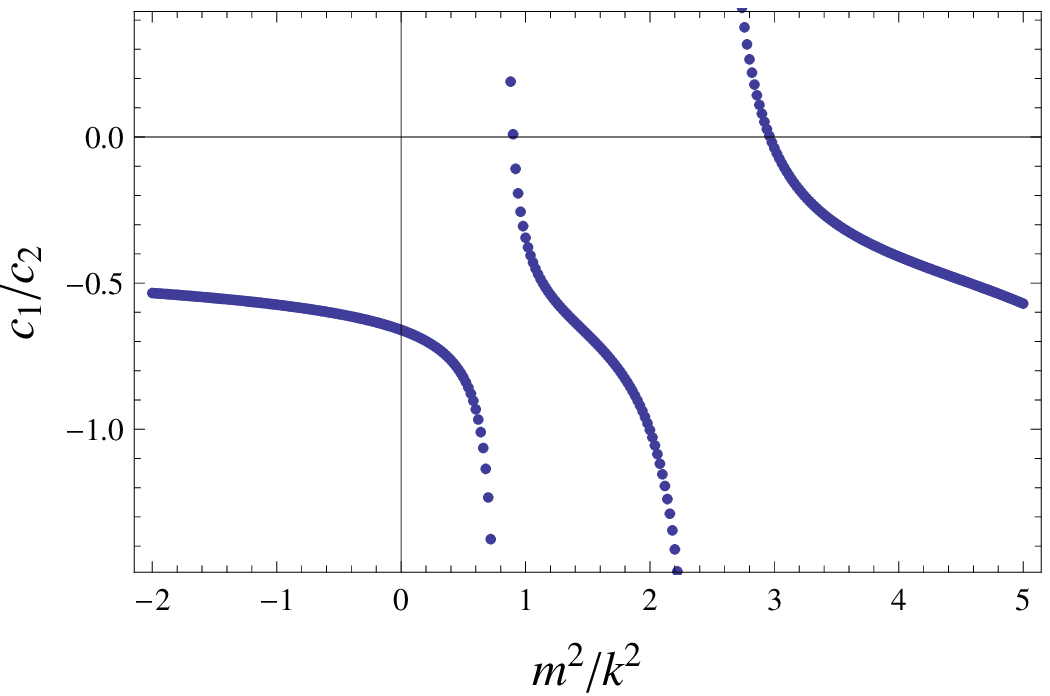,width=0.515\linewidth,clip=} \\
\end{tabular}

\caption{\it  $c_1/c_2$ as a  function  of  $m^2/k^2$   for $ky_s=1$ and $\nu=1.8$. Left panel: $a=3.6$. Right panel: $a=0.6$.
}
 \label{higgsfig}
 \end{figure}

Equation ~(\ref{higgsfluc}) can be brought into a Schr\"odinger form by defining $\tilde{\xi}=\xi \exp(-A)$.
The inner  inner product in this case becomes 
\be
(\tilde{\xi}_1,\tilde{\xi}_2)=\int_0^{y_s}e^{2A}  \tilde{\xi}_1^* \tilde{\xi}_2\, dy \,.
\label{normxit}
\ee
The dominant terms of Eq.~(\ref{higgsfluc}) near the singularity for $1<\nu<2$ are
\be \tilde{\xi}''(y)+
\frac{2
   \left(\nu^2-2\right) \tilde{\xi}(y)}{\nu^4
   (y-y_s)^2}=0,
\label{higgsasy}
\ee
and the asymptotic form of the solution is 
\be
\tilde{\xi}(y) = c_1(y_s-y)^{1-2/\nu^2}+c_2(y_s-y)^{2/\nu^2},
\ee
which is identical to the asymptotic behaviour of graviton fluctuations. As a result the $c_1/c_2$ plot in order to compute the KK spectrum can be done numerically. In Fig.~\ref{higgsfig} we give two examples of $c_1/c_2$ plots that correspond to qualitatively different situations. Both plots are made for $ky_s=1$, $\nu=1.8$ and $h_0=10^{-4}$. The value of $h_0$ is taken  small enough in order to have negligible backreaction to the metric. In the left panel of Fig.~\ref{higgsfig} the $c_1/c_2$ plot is computed with $a=3.6$. In this case we see that there is a tachyon-free regime for $0 \leq c_{12} \lesssim 0.2$. In the right panel of Fig.~\ref{higgsfig} for $a=0.6$ the tachyon-free regime is for $c_{12} \lesssim -0.65$ and $c_{12} \geq 0$.

\begin{figure}[t]
\begin{center}
 \epsfig{file=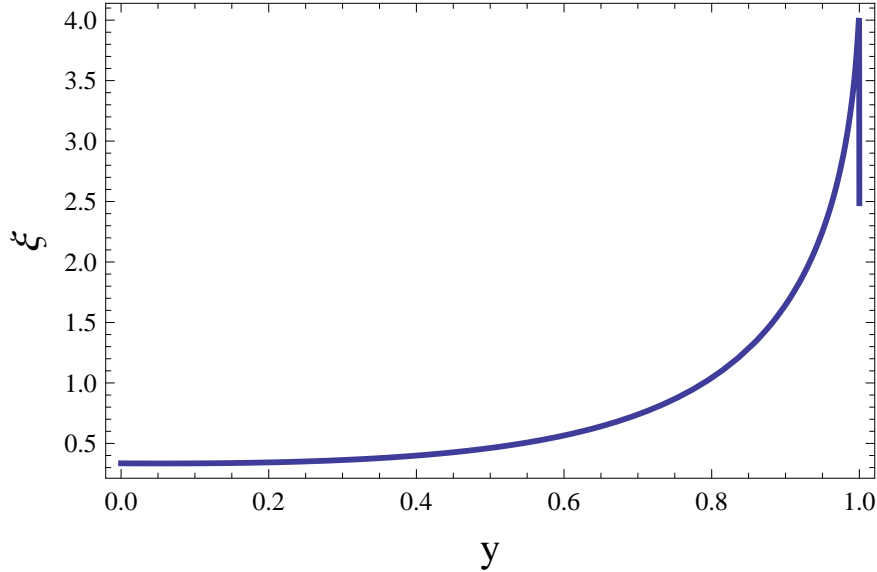,width=0.7\linewidth,clip=}
\end{center}
\caption{ \it Plot of a normalized mode:  $\xi$  as a function of $y$ for $\nu=1.8$, $ky_s=1$, $a=0.6$ and $m^2=0$.  }
 \label{profile}
 \end{figure}

Of course, the above  discussion for EWSB is far from complete. This is mainly due to the fact that we have three free parameters $a$, $\nu$ and $h_0$. It is quite a complex task to understand the behavior of the KK spectrum as we vary those three parameters independently. The plots in Fig.~\ref{higgsfig} represent typical cases where the KK spectrum of the Higgs can be ghost free.
  An additional shortcoming is that we are not trying, by choosing $ky_s=1$, to address the hierarchy problem which would require values $ky_s \sim 30$. The reason we did not use such large values of $ky_s $ is that the numerical algorithms we use to construct $c_1/c_2$ plots are not accurate enough in this case due to the large exponential factors which appear in Eq.~(\ref{higgsfluc})~\footnote{Nevertheless we have been able to check that for $ky_s \in \{0,8\}$ the qualitative structure of the $c_1/c_2$ plots remain unchanged.}. However, the main effect of large $ky_s$ is to scale down the $m^2$ values of the KK modes as a result of the warp factor. In fact we can see from Eq.~(\ref{normxi}) that $\xi$  provides a direct physical interpretation of localization properties along the extra dimension. The profile of normalized KK modes for $\xi$, see e.g.~the plot in Fig.~\ref{profile},
  is always localized near the singularity, say at $y_1=y_s-\ell$ for $k\ell\ll ky_s$. It turns out that in the support of $\xi$ the term $m^2e^{2A}\simeq m^2e^{2A(y_1)}$ and for values of $m^2e^{2A(y_1)}\sim k^2$, as the solution of Eq.~(\ref{higgsfluc}) requires all terms to have the same order of magnitude, the eigenvalues are warped down as $m\sim e^{-A(y_1)}k$ which can accommodate an electroweak Higgs from the Planckian value of $k$. Although our results for EWSB cannot be used directly for a phenomenological analysis, we believe that  they make up  a useful guide for future efforts towards a viable EWSB model.  

\section{Discussion and Outlook}
\label{conclusion}
The problem of specifying boundary conditions in soft-wall models at the position of the singularity
has been addressed in the past in the literature. When the warp factor has a logarithmic divergence it is possible to have  two independent normalizable  solutions for the graviton  eigenvalue equation. The approach followed in Refs.~\cite{Gursoy:2007er,Kiritsis:2006ua}
is that one has to fix the value of the ratio $c_1/c_2$, where $c_1$ and $c_2$ are the coefficients of the two independent graviton solutions, in order to be consistent with unitarity. For a specific value $c_1/c_2=0$  the lowest KK mode is massless. An equivalent proposal~\cite{Cabrer:2009we} is to assume the $e^{-4A(y_s)}h'(y_s)=0$ boundary condition. We have supplemented   previous works by finding that only the value $c_1/c_2=0$ is acceptable as every other choice 
is plagued by an instability of the background. For this particular value the lowest graviton KK mode is massless. We presented arguments showing that this result does not depend on the exact form of the warp factor, but rather on its asymptotic behaviour near the singularity. 

Along with the spectrum of gravitons we studied scalar and dilaton perturbations on the \emph{soft-wall} background. For the gauge choice we used the only dynamical scalar perturbation that corresponds to the radion. In this case one of the independent solutions of the eigenvalue problem is non-normalizable. It is common throughout the literature to exclude the non-normalizable solution from the spectrum. In analogy to the treatment of singular potentials in non-relativistic  quantum mechanics~\cite{Case,singular}, we adopted the point of view of not rejecting  the non-normalizable solution altogether, but assuming that it will become normalizable  when unknown physics related to the curvature singularity is taken into account. The resulting spectrum will strongly depend on the details of the resolution of the singularity. The most economical regularization is to introduce an IR brane near the position of the singularity. As an example we calculated the radion spectrum in this case finding that it is tachyon free for a wide range of boundary conditions. 

The approach developed in this work can be applied to the study of massive scalar, vector and spinor fields on \emph{soft-wall} backgrounds within studies beyond the Standard Model , using  extra dimensions. A particular application has been done in this paper to the case of the Standard Model Higgs propagating in the bulk of the singular metric defining the soft wall assuming EWSB. In this case it was understood that it is possible to have a Higgs KK  spectrum that is  free of tachyons for a range of $c_1/c_2$ values without introducing an ad hoc IR brane. The mass of the lowest KK mode depends on the value of the parameter  $c_1/c_2$. From the point of view of AdS/CFT it would be interesting to better understand the role of non-normalizable graviton solutions and to relate their regularization to nontrivial IR dynamics.

\section*{Acknowledgments}
The work of N. Brouzakis has been co-financed by the European Union (European Social Fund - ESF) and Greek national
funds through the Operational Program ``Education and Lifelong Learning" of the National Strategic Reference
Framework (NSRF) - Research Funding Program: ``THALIS. Investing in the society of knowledge through the
European Social Fund". The work of M.Quiros  has  been  supported by the Spanish Consolider-Ingenio 2010 Programme CPAN (CSD2007-00042) and by CICYT-FEDER-FPA2011-25948. We want to thank Joan Cabrer for having participated in the early stages of this paper.

\appendix

\section{Appendix}
\label{appendix}
In this appendix we give a brief review of some facts regarding unbounded differential operators and their spectrum. For a more precise exposition the reader can consult Refs.~\cite{Reed:1975uy} and  \cite{Stone:2009}. We will also argue that fixing $c_1/c_2$ is sufficient to ensure that the evolution is unitary. 

Our goal is to study the mass spectrum described by Eq.~(\ref{schro}). The corresponding operator is 
\be
\Op_s=-e^{-2A(y)} \frac{d^2}{dy^2}+e^{-2A(y)} V(y),
\ee
where $V(y)$ is given by Eq.~(\ref{pot}). This operator acts on a subset of the Hilbert space $L^2[0,y_s)$ with  weight function $w(x)=e^{2A(y)}$. Thus the norm on this Hilbert space is defined by
\be
(f,g)=\int_0^{y_s} e^{2A(y)} f^*(y)g(y)dy.
\ee
The asymptotic form of eigenfunctions of $\Op_s$ is given by Eq.~(\ref{asymp}). We assume that the domain of the operator $\mathcal{D}(\Op_s)$ consists of functions with this asymptotic behaviour. An operator $\Op_s$ is called self-adjoint if:
\begin{itemize}
\item[1.] It is symmetric when acting on functions $f,g \in \mathcal{D}(\Op_s)$,
\be
(f,\Op_s g)=(\Op_s f , g),
\label{symmetric}
\ee
\item[2.] The domain of the adjoint operator $\mathcal{D}(\Op_s^*)\equiv\mathcal{D}(\Op_s)$ .
\end{itemize}

The domain of the adjoint is defined to consist of \emph{all} functions $f \in  L^2[0,y_s)$ that 
satisfy  Eq.~(\ref{symmetric}) for every $g \in\mathcal{D}(\Op_s)$. Notice that, even if $\Op_s^*$ is formally identical to $\Op_s$ in reality they can be distinct since it is possible that they act on different domains. The second requirement is needed  to ensure that the operator is symmetric also when it acts on functions of the form 
\be
 \psi(t)=e^{it\Op_s}\psi(0) \qquad \psi(0) \in  \mathcal{D}(\Op_s),
\ee
that will be generated by the time evolution.
 
If $\Op_s$ is self-adjoint then it  has the following properties:
\begin{itemize}
\item A spectrum of real eigenvalues.
\item Eigenfunctions that  form an orthonormal basis which is complete in $L^2[0,y_s)$.
\item It generates unitary evolution.
\end{itemize}
Let us consider $f,g$ to be functions that satisfy the boundary condition of Eq.~(\ref{boundary}) and 
having an asymptotic behaviour near the singularity of the form of Eq.~(\ref{asymp})
\bea
f(y) &\sim& a_1(y_s-y)^{1-2/\nu^2}+a_2(y_s-y)^{2/\nu^2} ,\\
g(y) &\sim& b_1(y_s-y)^{1-2/\nu^2}+b_2(y_s-y)^{2/\nu^2}  .
\eea
Now we can check whether $\Op_s$ is symmetric with respect to $f,g$ or not. Performing two consecutive  integrations by parts in the interval $[0,y_s)$, we can see that 
\be
(f, \Op_s g)=(\Op_s f, g)+\left[f^*(y)g'(y)-f'^*(y)g(y)\right]^{y \to y_s}_0 .
\ee
The boundary term at $y=0$ vanishes  due to the boundary condition of Eq.~(\ref{boundary}). Taking into account the asymptotic form of $f,g$ we see that 
\be
(f, \Op_s g)=(\Op_s f, g)+\left(a_1^*b_2-a_2^*b_1 \right) \left(1-\frac{4}{\nu^2}\right).
\ee
The operator $\Op_s$ is then symmetric provided that $a_1,a_2 \in \mathbb{R}$ and 
\be
\frac{a_1}{a_2}=\frac{b_1}{b_2}\equiv\frac{c_1}{c_2}.
\ee
We have  proven that if we choose a  $ \mathcal{D}(\Op_s)$ that consists of functions with fixed $c_1/c_2$, the operator is symmetric and the domain of the adjoint is identical with the domain of $\Op_s$: in this case the operator is self-adjoint. It is possible to see that time evolution with  $O_s$ will not change the asymptotic behaviour and the value of $c_1/c_2$. 

Another mathematically  rigorous proof of the $c_1/c_2$ statement can be given using von Neumann's theory of self-adjoint extensions (see~\cite{Reed:1975uy}). One can start assuming 
that $\mathcal{D}(\Op_s)$ consists of functions satisfying Eq.~(\ref{boundary}) and having  a compact support at the singularity. The deficiency indices in this case are 
\be
n_{\pm}=\dim(Ker(\Op_s \pm i))=1,
\ee
Thus we can have a self-adjoint extension of $\Op_s$ by adding to the domain of $\Op_s$ functions of the form 
\be 
\psi_e=\psi_++e^{i \gamma} \psi_-,
\ee
where $\psi_{\pm}$ satisfy  $\Op_s \psi_{\pm}=\pm i \psi_{\pm}$ and $\gamma$ is a fixed number in $\mathbb{R}$. Since $\psi_{\pm}$ are eigenfunctions of $\Op_s$ they have the usual asymptotic behaviour mentioned above:
\bea
\psi_+&=& c_{1+}(y_s-y)^{1-2/\nu^2}+c_{2+}(y_s-y)^{2/\nu^2} \\
\psi_-&=& c_{1-}(y_s-y)^{1-2/\nu^2}+c_{2-}(y_s-y)^{2/\nu^2}.
\eea
Since $\psi_{\pm}$ are unique solutions, $\{c_{1+},c_{1-},c_{2+},c_{2-}\}$  are fixed numbers satisfying $c_{1,2+}^*=c_{1,2-}$. The asymptotic form of the functions $\psi_e$ that extend the domain of the operator is
\be
\psi_e \sim \left(c_{1+}+e^{i \gamma} c_{1-}\right) (y_s-y)^{1-2/\nu^2}+
\left(c_{2+}+e^{i \gamma}c_{2-}\right)(y_s-y)^{2/\nu^2}.
\ee
 It is trivial to check that the ratio
\be 
\frac{c_{1+}+e^{i \gamma} c_{1-}}{ c_{2+}+e^{i \gamma}c_{2-}} \equiv \frac{c_1}{c_2},
\ee
is an arbitrary real number parametrized by the angle $\gamma$.


\begin{thebibliography}{99}

\bibitem{Randall:1999ee}
  L.~Randall and R.~Sundrum,
  ``A Large mass hierarchy from a small extra dimension,''
  Phys.\ Rev.\ Lett.\  {\bf 83} (1999) 3370
  [hep-ph/9905221].


\bibitem{Maldacena:1997re} 
  J.~M.~Maldacena,
  ``The Large N limit of superconformal field theories and supergravity,''
  Adv.\ Theor.\ Math.\ Phys.\  {\bf 2}, 231 (1998)
  [hep-th/9711200].
\bibitem{Gubser:1998bc} 
  S.~S.~Gubser, I.~R.~Klebanov and A.~M.~Polyakov,
  ``Gauge theory correlators from noncritical string theory,''
  Phys.\ Lett.\ B {\bf 428}, 105 (1998)
  [hep-th/9802109].

\bibitem{Erlich:2005qh} 
  J.~Erlich, E.~Katz, D.~T.~Son and M.~A.~Stephanov,
  ``QCD and a holographic model of hadrons,''
  Phys.\ Rev.\ Lett.\  {\bf 95}, 261602 (2005)
  [hep-ph/0501128].

\bibitem{Csaki:2006ji} 
  C.~Csaki and M.~Reece,
  ``Toward a systematic holographic QCD: A Braneless approach,''
  JHEP {\bf 0705}, 062 (2007)
  [hep-ph/0608266];
  L.~Da Rold and A.~Pomarol,
  ``Chiral symmetry breaking from five dimensional spaces,''
  Nucl.\ Phys.\ B {\bf 721}, 79 (2005)
  [hep-ph/0501218].



\bibitem{Goldberger:1999uk} 
  W.~D.~Goldberger and M.~B.~Wise,
  ``Modulus stabilization with bulk fields,''
  Phys.\ Rev.\ Lett.\  {\bf 83}, 4922 (1999)
  [hep-ph/9907447].

\bibitem{Kachru:2000hf} 
  S.~Kachru, M.~B.~Schulz and E.~Silverstein,
  ``Selftuning flat domain walls in 5-D gravity and string theory,''
  Phys.\ Rev.\ D {\bf 62}, 045021 (2000)
  [hep-th/0001206];
  N.~Arkani-Hamed, S.~Dimopoulos, N.~Kaloper and R.~Sundrum,
  ``A Small cosmological constant from a large extra dimension,''
  Phys.\ Lett.\ B {\bf 480}, 193 (2000)
  [hep-th/0001197].

\bibitem{Forste:2000ps} 
  S.~Forste, Z.~Lalak, S.~Lavignac and H.~P.~Nilles,
  ``A Comment on selftuning and vanishing cosmological constant in the brane world,''
  Phys.\ Lett.\ B {\bf 481}, 360 (2000)
  [hep-th/0002164].

\bibitem{Karch:2006pv} 
  A.~Karch, E.~Katz, D.~T.~Son and M.~A.~Stephanov,
  ``Linear confinement and AdS/QCD,''
  Phys.\ Rev.\ D {\bf 74}, 015005 (2006)
  [hep-ph/0602229].

\bibitem{Gursoy:2007er} 
  U.~Gursoy, E.~Kiritsis and F.~Nitti,
  ``Exploring improved holographic theories for QCD: Part II,''
  JHEP {\bf 0802}, 019 (2008)
  [arXiv:0707.1349 [hep-th]].

\bibitem{Batell:2008zm} 
  B.~Batell and T.~Gherghetta,
  ``Dynamical Soft-Wall AdS/QCD,''
  Phys.\ Rev.\ D {\bf 78}, 026002 (2008)
  [arXiv:0801.4383 [hep-ph]].



\bibitem{bulkfields} 
  H.~Davoudiasl, J.~L.~Hewett and T.~G.~Rizzo,
  ``Bulk gauge fields in the Randall-Sundrum model,''
  Phys.\ Lett.\ B {\bf 473}, 43 (2000)
  [hep-ph/9911262];
  A.~Pomarol,
  ``Gauge bosons in a five-dimensional theory with localized gravity,''
  Phys.\ Lett.\ B {\bf 486}, 153 (2000)
  [hep-ph/9911294]
; Y.~Grossman and M.~Neubert,
  ``Neutrino masses and mixings in nonfactorizable geometry,''
  Phys.\ Lett.\ B {\bf 474}, 361 (2000)
  [hep-ph/9912408];
  S.~Chang, J.~Hisano, H.~Nakano, N.~Okada and M.~Yamaguchi,
  ``Bulk standard model in the Randall-Sundrum background,''
  Phys.\ Rev.\ D {\bf 62}, 084025 (2000)
  [hep-ph/9912498];
  T.~Gherghetta and A.~Pomarol,
  ``Bulk fields and supersymmetry in a slice of AdS,''
  Nucl.\ Phys.\ B {\bf 586}, 141 (2000)
  [hep-ph/0003129].



\bibitem{Contino:2010rs} 
  R.~Contino,
  ``The Higgs as a Composite Nambu-Goldstone Boson,''
  arXiv:1005.4269 [hep-ph].

\bibitem{Gherghetta:2010cj} 
  T.~Gherghetta,
  ``TASI Lectures on a Holographic View of Beyond the Standard Model Physics,''
  arXiv:1008.2570 [hep-ph].

\bibitem{Huber:2000fh} 
  S.~J.~Huber and Q.~Shafi,
  ``Higgs mechanism and bulk gauge boson masses in the Randall-Sundrum model,''
  Phys.\ Rev.\ D {\bf 63}, 045010 (2001)
  [hep-ph/0005286].

\bibitem{Falkowski:2008fz}
  A.~Falkowski and M.~Perez-Victoria,
  ``Electroweak Breaking on a Soft Wall,''
  JHEP {\bf 0812} (2008) 107
  [arXiv:0806.1737 [hep-ph]].

\bibitem{Batell:2008me} 
  B.~Batell, T.~Gherghetta and D.~Sword,
  ``The Soft-Wall Standard Model,''
  Phys.\ Rev.\ D {\bf 78}, 116011 (2008)
  [arXiv:0808.3977 [hep-ph]].

\bibitem{Cabrer:2011fb} 
  J.~A.~Cabrer, G.~von Gersdorff and M.~Quiros,
  ``Suppressing Electroweak Precision Observables in 5D Warped Models,''
  JHEP {\bf 1105}, 083 (2011)
  [arXiv:1103.1388 [hep-ph]].

\bibitem{Carmona:2011ib}
  A.~Carmona, E.~Ponton and J.~Santiago,
  ``Phenomenology of Non-Custodial Warped Models,''
  JHEP {\bf 1110}, 137 (2011) [arXiv:1107.1500 [hep-ph]].

\bibitem{deBlas:2012qf}
  J.~de Blas, A.~Delgado, B.~Ostdiek and A.~de la Puente,
  ``LHC Signals of Non-Custodial Warped 5D Models,''
  Phys.\ Rev.\ D {\bf 86}, 015028 (2012) [arXiv:1206.0699 [hep-ph]].

\bibitem{singular} 
  W.~Frank, D.~J.~Land and R.~M.~Spector,
  ``Singular potentials,''
  Rev.\ Mod.\ Phys.\  {\bf 43}, 36 (1971).

\bibitem{Case}
  K.M.Case,
 ``Singular Potentials,''
  Phys.\ Rev.\   {\bf 80}, 797-806 (1950).

\bibitem{barry}
M.~Reed and B.~ Simon, ``Methods of Modern Mathematical Physics'' (Academic
Press 1980).


\bibitem{Wald:1980jn}
  R.~M.~Wald,
  ``Dynamics In Nonglobally Hyperbolic, Static Space-Times,''
  J.\ Math.\ Phys.\  {\bf 21}, 2802 (1980);
  A.~Ishibashi and R.~M.~Wald,
  ``Dynamics in non-globally-hyperbolic static spacetimes. II: General
  analysis of prescriptions for dynamics,''
  Class.\ Quant.\ Grav.\  {\bf 20}, 3815 (2003)
  [arXiv:gr-qc/0305012];
  A.~Ishibashi and R.~M.~Wald,
  ``Dynamics in non-globally hyperbolic static spacetimes. III: anti-de  Sitter
  spacetime,''
  Class.\ Quant.\ Grav.\  {\bf 21}, 2981 (2004)
  [arXiv:hep-th/0402184].

\bibitem{Brax:2001cx}
  P.~Brax and A.~C.~Davis,
  ``On brane cosmology and naked singularities,''
  Phys.\ Lett.\  B {\bf 513}, 156 (2001)
  [arXiv:hep-th/0105269].

\bibitem{Horowitz:1995gi}
  G.~T.~Horowitz and D.~Marolf,
  ``Quantum Probes Of Space-Time Singularities,''
  Phys.\ Rev.\  D {\bf 52}, 5670 (1995)
  [arXiv:gr-qc/9504028].


\bibitem{Cabrer:2009we} 
  J.~A.~Cabrer, G.~von Gersdorff and M.~Quiros,
  ``Soft-Wall Stabilization,''
  New J.\ Phys.\  {\bf 12}, 075012 (2010)
  [arXiv:0907.5361 [hep-ph]].

\bibitem{DeWolfe:1999cp} 
  O.~DeWolfe, D.~Z.~Freedman, S.~S.~Gubser and A.~Karch,
  ``Modeling the fifth-dimension with scalars and gravity,''
  Phys.\ Rev.\ D {\bf 62}, 046008 (2000)
  [hep-th/9909134].


\bibitem{Gubser:2000nd} 
  S.~S.~Gubser,
  ``Curvature singularities: The Good, the bad, and the naked,''
  Adv.\ Theor.\ Math.\ Phys.\  {\bf 4}, 679 (2000)
  [hep-th/0002160].

\bibitem{Kim:2000yq} 
  H.~D.~Kim,
  ``A Criterion for admissible singularities in brane world,''
  Phys.\ Rev.\ D {\bf 63}, 124001 (2001)
  [hep-th/0012091].

\bibitem{Cabrer:2010si} 
  J.~A.~Cabrer, G.~von Gersdorff and M.~Quiros,
  ``Warped Electroweak Breaking Without Custodial Symmetry,''
  Phys.\ Lett.\ B {\bf 697}, 208 (2011)
  [arXiv:1011.2205 [hep-ph]].

\bibitem{Csaki:2000zn} 
  C.~Csaki, M.~L.~Graesser and G.~D.~Kribs,
 ``Radion dynamics and electroweak physics,''
  Phys.\ Rev.\ D {\bf 63}, 065002 (2001)
  [hep-th/0008151].

\bibitem{Kiritsis:2006ua} 
  E.~Kiritsis and F.~Nitti,
  ``On massless 4D gravitons from asymptotically AdS(5) space-times,''
  Nucl.\ Phys.\ B {\bf 772}, 67 (2007)
  [hep-th/0611344].

\bibitem{Reed:1975uy} 
  M.~Reed and B.~Simon,
  ``Methods of Modern Mathematical Physics. 2. Fourier Analysis, Self-Adjointness,''
 (Academic Press, New York, 1975), 361p.

\bibitem{Stone:2009} 
  M.~Stone and P.~Goldbart,
  ``Mathematics for Physics,''
  (Cambridge University Press, New York, 2009), 806p.
 
\end{thebibliography}
\end{document}